\documentclass[runningheads]{llncs}
\usepackage{amsfonts}

\usepackage{latexsym}
\usepackage{amssymb}
\usepackage{epsfig}
\usepackage{amsmath}
\usepackage[mathscr]{eucal}
\usepackage{subfigure}
\usepackage{graphicx}

\renewcommand{\thefigure}{\arabic{figure}}
\begin{document}

\title{Probabilistic Bounds for Data Storage with Feature Selection and Undersampling}
\titlerunning{Data Storage Bound}

\author{ \textbf{Ghurumuruhan Ganesan}}
\authorrunning{G. Ganesan}
\institute{IISER Bhopal\\
\email{gganesan82@gmail.com }}

\date{}
\maketitle

\begin{abstract}
In this paper we consider data storage from a probabilistic point of view and obtain bounds for efficient storage in the presence of feature selection and undersampling, both of which are important from the data science perspective. First, we consider encoding of correlated sources for nonstationary data and obtain a Slepian-Wolf type result for the probability of error. We then reinterpret our result by allowing one source to be the set of features to be discarded and other source to be remaining data to be encoded. Next, we consider neighbourhood domination in random graphs where we impose the condition that a fraction of neighbourhood must be present for each vertex and obtain optimal bounds on the minimum size of such a set. We show how such sets are useful for data undersampling in the presence of imbalanced datasets and briefly illustrate our result using~\(k-\)nearest neighbours type classification rules as an example.

\vspace{0.1in} \noindent \textbf{Key words:} Data Storage, Probabilistic Bounds, Correlated Encoding, Feature Selection, Neighbourhood Domination, Data Undersampling.

\vspace{0.1in} \noindent \textbf{AMS 2000 Subject Classification:} Primary:
94A15, 94A24.
\end{abstract}

\bigskip

\setcounter{equation}{0}
\renewcommand\theequation{\arabic{section}.\arabic{equation}}
\section{Introduction} \label{intro}
Data feature selection and undersampling are important topics from both theoretical and practical perspectives and in this paper, we use a probabilistic approach to obtain theoretical bounds for data storage after feature selection and/or undersampling. Feature selection broadly fall into two categories: supervised and unsupervised. In supervised feature selection, the target variable is used as a guide in deciding which features of the data to retain and which features to discard. On the other hand, as the name suggests, unsupervised feature selection is done without using the target variable. There are many statistical methods available for both supervised and unsupervised feature selection and for more details, we refer to Chapter~\(19\) of~\cite{kuhn}.

We are interested in the problem of efficient data storage \emph{after} feature selection using Slepian-Wolf coding.  Apart from traditional source coding, Slepian-Wolf coding has also been extensively used in wireless networks. For example~\cite{pradhan} described distributed compression techniques for dense microsensor networks using syndromes and later~\cite{hu} proposed Slepian-Wolf cooperation to improve inter-user outage performance. In a related work~\cite{zheng} used Slepian-Wolf encoding for data aggregation in cluster-based wireless networks and~\cite{timo} studied the performance of broadcasting with base-station cooperation and Slepian-Wolf coding.

In this paper, we obtain a Slepian-Wolf type encoding result for non-stationary data and illustrate how information from the discarded features could be used to store the remaining data in a more efficient manner.

In the next part of our paper, we study neighbourhood domination in random graphs with applications to data undersampling. Graph domination is important from both theoretical and application perspectives and many variants of graph domination have been studied in different contexts. For domination in random graphs,~\cite{wieland} obtains two point concentration for the domination number of~\(G\) when~\(p\) is essentially a constant and then extended for a wider range of~\(p\) in~\cite{glebov}. Since then many other variants of domination have also been studied (see for e.g.~\cite{clark}\cite{wang}) and recently,~\cite{ganesan} studies robustness of dominating sets when a small deterministic set of edges is removed from the parent complete graph.

In this paper, we study neighbourhood domination in random graphs where we impose the condition that a fraction of the neighbourhood of each vertex is present in the resulting dominating set. We use the probabilistic method to obtain sufficient conditions for the existence of neighbourhood dominating sets of minimum possible size and demonstrate optimality of our estimates by obtaining a lower bound of the same order, for homogenous random graphs with common edge probability~\(p.\) We also briefly illustrate a data undersampling methodology using neighbourhood domination and explain its applications to imbalanced learning.

The paper is organized as follows: In Section~\ref{sec_dep_code}, we state and prove our first result regarding a Slepian-Wolf type bound for non-stationary data. As a consequence, we also demonstrate how savings in data storage could be achieved by using information from the discarded features. Next in Section~\ref{sec_neigh}, we describe our second result involving the minimum size of neighbourhood dominating sets in random graphs. We also illustrate an application of our methodology in constrained undersampling of imbalanced data. Finally in Section~\ref{sec_conc}, we state our conclusion and propose potential future directions.


\setcounter{equation}{0}
\renewcommand\theequation{\arabic{section}.\arabic{equation}}
\section{Dependent Encoding based on Feature Selection}\label{sec_dep_code}
Let~\({\cal X}\) and~\({\cal Y}\) be arbitrary finite sets and for~\(1 \leq i \leq n,\) let~\((X_i,Y_i) \in {\cal X} \times {\cal Y}\) be a random element with distribution~\(\mathbb{P}(X_i=a,Y_i=b) = p_i(a,b)\) and corresponding marginals~\[p_{i,X}(.) := \sum_{b \in {\cal Y}} p_i(.,b)\text{ and }p_{i,Y}(.) := \sum_{a \in {\cal X}} p(x,.).\] The tuples~\((X_i,Y_i), 1 \leq i \leq n \) are independent but not necessarily identically distributed.

An~\(n-\)length binary code of rate~\(R\) is a deterministic set~\({\cal C} \subset \{0,1\}^{n}\) of size~\(2^{nR}.\)
\begin{definition}\label{def_one} An~\(n-\)length,~\(Y-\)dependent encoder based on~\({\cal C}\) is a set of one-to-one maps~\(f :=\{f_y(.)\}_{y \in {\cal Y}^{n}}\) such that~\(f_y : {\cal C} \rightarrow {\cal X}^{n}\) for each~\(y.\) We  define the probability of error for the encoding scheme~\((f,{\cal C})\) as
\begin{equation}\label{dec_err2}
q(f,{\cal C}) :=  \mathbb{P}\left(U \notin f_V({\cal C})\right),
\end{equation}
where~\(U := (X_1,\ldots,X_n)\) and~\(V := (Y_1,\ldots,Y_n).\)
\end{definition}
In the context of data analysis,~\((X_i,Y_i)\) represents the~\(i^{th}\) data point where~\(X_i\) represents the part of the data to be encoded and~\(Y_i\) is the part of the data that is to be discarded based on feature selection. The goal is to use the information from~\((Y_1,\ldots,Y_n)\) to encode~\((X_1,\ldots,X_n)\) as efficiently as possible and we explain this in more detail at the end of this section.

We also remark that we do not assume that~\((X_i,Y_i)\) are identically distributed to allow for the possibility of measurement errors in data due to inherent statistical noise and/or human errors.


Defining
\begin{equation}\label{ent_def}
H(X_i,Y_i) := -\sum_{x\in{\cal X},y\in{\cal Y}}p_i(x,y)\log{p_i(x,y)} \text{ and } H(Y_i) := -\sum_{y \in {\cal Y}}p_{i,Y}(y)\log{p_{i,Y}(y)}
\end{equation} to be the respective entropies of~\((X_i,Y_i)\) and~\(Y_i,\) we have the following result. Throughout, logarithms are to the base~\(2\) and constants do not depend on~\(n.\)
\begin{theorem}\label{main_thm} Suppose
\begin{equation}\label{prob_cond}
\epsilon_1 \leq \min_{i,x,y}p_{i}(x,y) \leq \max_{i,x,y} p_{i}(x,y) \leq \epsilon_2,\;\;\;\epsilon_1 \leq \min_{i,y}p_{i,Y}(y) \leq \max_{i,y} p_{i,Y}(y) \leq \epsilon_2
\end{equation}
and
\begin{equation}\label{ent_cond}
\frac{1}{n}\sum_{i=1}^{n}H(X_i,Y_i) \longrightarrow H_{XY} \text{ and } \frac{1}{n}\sum_{i=1}^{n}H(Y_i) \longrightarrow H_Y
\end{equation}
for some positive finite constants~\(\epsilon_1, \epsilon_2, H_{XY}\) and~\(H_Y.\) Let~\(\epsilon > 0\) be an arbitrary constant and let~\({\cal C}\) be any deterministic~\(n-\)length binary code of size~\(2^{nR}\) for some constant~\(0 < R < 1.\)\\
\((a)\) If~\(R > H_{XY}-H_Y+5\epsilon\)  then there exists an encoding scheme~\((f_0,{\cal C})\) such that the error probability
\begin{equation}\label{err_up}
q(f_0,{\cal C}) \leq 4\epsilon.
\end{equation}
\((b)\) If~\(R < H_{XY}-H_Y-3\epsilon\)  then for any encoding scheme~\((f,{\cal C})\) we have
\begin{equation}\label{err_low}
q(f,{\cal C}) \geq 1-4\epsilon.
\end{equation}
\end{theorem}
If the~\((X_i,Y_i)\) are independent and identically distributed (i.i.d.) then~\(H_{XY} = H(X,Y)\) and~\(H_Y = H(Y)\) are entropies as in~(\ref{ent_def}) and so~\(H_{XY}-X_Y = H(X \mid Y)\) is the conditional entropy of~\(X\) given~\(Y.\) This is the usual Slepian-Wolf encoding bound for correlated sources and for more details in this setting, we refer to Chapter~\(15,\) pp.~\(549\) of~\cite{cover}.

In general, for non-i.i.d.\ distributions, Theorem~\ref{main_thm} obtains a sharp threshold for the probability of encoding error for rates above and below  the threshold value of~\(H_{XY}-H_Y.\)



\emph{Proof of Theorem~\ref{main_thm}}: We begin with some preliminary computations regarding the size of certain typical sets used in our encoding process.
Let~\(U = (X_1,\ldots,X_n), V = (Y_1,\ldots,Y_n)\) where~\(\{(X_i,Y_i)\}\) are as described prior to~(\ref{dec_err2}) and let~\(p_{UV}(.,.)\) be the distribution of~\((X,Y)\) with corresponding marginals~\[p_U(.) := \sum_{y \in {\cal Y}^{n}}p_{UV}(,y) \text{ and }p_V(.) := \sum_{x \in {\cal X}^{n}} p_{UV}(x,.).\] 

For~\(\epsilon >0\) arbitrary, we define the typical set
\begin{equation}\label{ey_def}
E_Y := \{y \in {\cal Y}^{n}  : 2^{-n(H_Y+\epsilon)} \leq p_V(y) \leq 2^{-n(H_Y-\epsilon)}\}
\end{equation}
where~\(H_Y\) is as in~(\ref{ent_cond}).  To estimate the probability of the event~\(E_Y\) we first use the fact that~\(\{Y_i\}\) are independent to get that
\begin{equation}\label{thmulp}
\mathbb{E}\left(-\frac{1}{n}\sum_{i=1}^{n}(\log{p_{i,Y}(Y_i)}-H(Y_i))\right)^2 = \frac{1}{n^2} \sum_{i=1}^{n} \mathbb{E}(\log{p_{i,Y}(Y_i)}-H(Y_i))^2.
\end{equation}
Now from~(\ref{prob_cond}), we know that~\(\mathbb{E}(\log{p_{i,Y}(Y_i)}-H(Y_i))^2 \leq C\) for some constant\\\(C > 0\) and so
\begin{equation}\label{entrop_mean}
\mathbb{E}\left(-\frac{1}{n}\sum_{i=1}^{n}(\log{p_{i,Y}(Y_i)}-H(Y_i))\right)^2 \leq \frac{C}{n} \longrightarrow 0
\end{equation}
as~\(n \rightarrow \infty.\) Combining~(\ref{entrop_mean}) with~(\ref{ent_cond}), we then get that
\[\mathbb{E}\left(-\frac{1}{n}\log{p_V(V)}-H_Y\right)^2 \longrightarrow 0\]  as~\(n \rightarrow \infty.\) Thus~\(-\frac{1}{n}\log{p_V(V)} \longrightarrow H_Y\) in probability and so
\begin{equation}\label{ey_prob}
\mathbb{P}(E_Y) \geq 1-\epsilon
\end{equation}
for all~\(n\) large.

Similarly defining
\begin{equation}\label{exy_def}
E_{XY} := \{(x,y) \in {\cal X}^{n} \times {\cal Y}^{n}  : 2^{-n(H_{XY}+\epsilon)} \leq p_{UV}(x,y) \leq 2^{-n(H_{XY}-\epsilon)}\}
\end{equation}
and performing an analogous analysis as above, we have for all~\(n\) large that\\\(\mathbb{P}(E_{XY}) \geq 1-\epsilon.\)  From~(\ref{ey_prob}) and the union bound, we therefore get that
\begin{equation}\label{e_tot_est2}
\mathbb{P}(E_{XY} \cap E_Y) \geq 1-2\epsilon.
\end{equation}

If~\(\#E_Y\) denotes the cardinality of~\(E_Y,\) then by definition
\[ 1 \geq \mathbb{P}(E_Y) =\sum_{y \in E_Y} p_V(y) \geq \#E_Y \cdot \frac{1}{2^{n(H_Y+\epsilon)}}\] and so~\(\#E_Y \leq 2^{n(H_Y+\epsilon)}.\)
Similarly using~(\ref{ey_prob}) we also have that
\[1-\epsilon \leq\mathbb{P}(E_Y)  = \sum_{y \in E_Y} p_V(y) \leq \#E_Y \cdot \frac{1}{2^{n(H_Y-\epsilon)}}\] and so~\(\#E_Y \geq (1-\epsilon) 2^{n(H_Y-\epsilon)}.\) Combining we get
\begin{equation}\label{ey_size_est}
(1-\epsilon)2^{n(H_Y-\epsilon)} \leq \#E_{Y} \leq 2^{n(H_Y+\epsilon)}
\end{equation}
and arguing similarly we also get that
\begin{equation}\label{exy_size_est}
(1-\epsilon)2^{n(H_{XY}-\epsilon)} \leq \#E_{XY} \leq 2^{n(H_{XY}+\epsilon)}.
\end{equation}
This completes the preliminary computations part of our proof.


For obtaining Theorem~\ref{main_thm}\((a),\) we argue as follows. For~\(y \in E_Y\) define
\begin{equation}\label{ay_def}
A_y := \{x \in {\cal X}^{n} : (x,y) \in E_{XY}\}
\end{equation}
to be the slice of~\(E_{XY}\) for a given~\(y.\) If~\[B_Y :=\{ y \in E_Y : \#A_y \geq 2^{n(H_{XY}-H_Y + 5\epsilon)}\},\] then necessarily~\(\#B_Y \leq 2^{n(H_Y -3\epsilon)};\) for otherwise, we would get
\[\#E_{XY} \geq \#B_Y \cdot 2^{n(H_{XY}-H_Y+5\epsilon)} \geq 2^{n(H_{XY} + 2\epsilon)}\] and this contradicts~(\ref{exy_size_est}). Thus~\(\#B_Y \leq 2^{n(H_Y-3\epsilon)}\) and so from the definition of~\(E_Y\) in~(\ref{ey_def}) we get that \[\mathbb{P}(B_Y) \leq 2^{n(H_Y-3\epsilon)} \cdot 2^{-n(H_Y-\epsilon)} = \frac{1}{2^{2n\epsilon}}.\] Since~\(B_Y \subset E_Y,\) we get from~(\ref{ey_prob}) that
\begin{equation}\label{ey_by_est}
\mathbb{P}(E_Y \setminus B_Y) \geq  1-\epsilon - \frac{1}{2^{2n\epsilon}}.
\end{equation}

For each~\( y \in E_Y \setminus B_Y,\)  we have that~\(\#A_y \leq 2^{n(H_{XY}-H_Y + 5\epsilon)}\) and so letting~\({\cal C}\) be any~\(n-\)length binary code of size~\(2^{n(H_{XY}-H_Y+5\epsilon)}\) we define a one-to-one map\\\(f_y : {\cal C}  \rightarrow  A_y,\) for each~\(y \in E_Y \setminus B_Y.\) If~\((U,V)\) is the random tuple as described in the first paragraph of this proof, then from~(\ref{e_tot_est2}),~(\ref{ey_by_est}) and the union bound, we see that the probability of error~\(q(f,{\cal C})\) is at most \[\mathbb{P}((U,V) \notin E_{XY}) + \mathbb{P}(V \notin E_Y \setminus B_Y) \leq 3\epsilon + \frac{1}{2^{2n\epsilon}} \leq 4\epsilon\]
for all~\(n\) large.  This obtains the upper bound~(\ref{err_up}) and therefore completes the proof of part~\((a).\)

For the lower bound~(\ref{err_low}) in part~\((b),\) we argue as follows. Again let~\((U,V)\) be the random tuple as described in the first paragraph of this proof and for~\(y \in {\cal Y}^{n},\) let~\(A_y\) be the ``slice" of the set~\(E_{XY}\) as defined in~(\ref{ay_def}). We have that
\begin{equation}\label{av_low}
\mathbb{P}(A_V \cap E_Y) = \mathbb{P}\left(\{(U,V) \in E_{XY}\} \cap \{V \in E_Y\}\right) \geq 1-2\epsilon,
\end{equation}
as seen from~(\ref{e_tot_est2}). Expanding in terms of the elements of~\(E_Y,\) we also get
\[\mathbb{P}(A_V \cap E_Y) =  \sum_{y \in E_Y} \mathbb{P}_y(A_y)p_V(y)\]
where
\begin{equation}\label{py_def}
\mathbb{P}_y(A_y) := \mathbb{P}(A_V \mid V=y) = \sum_{x \in A_y} p_{U \mid V}(x \mid y)
\end{equation}  and~\(p_{U \mid V}(x \mid y) := \frac{p_{UV}(x,y)}{p_V(y)}.\)  Therefore defining \[C_Y := \{y \in E_Y : \mathbb{P}_y(A_y) \geq \epsilon\} \]  we also get the upper bound
\begin{eqnarray}\label{av_up}
\mathbb{P}(A_V \cap E_Y) &\leq&  \sum_{y \in C_Y} p_V(y) + \epsilon \sum_{y \in E_Y \setminus C_Y} p_V(y) \nonumber\\
&\leq& \mathbb{P}(C_Y) + \epsilon.
\end{eqnarray}
Combining~(\ref{av_low}) and~(\ref{av_up}), we get that
\begin{equation}\label{cy_est}
\mathbb{P}(C_Y) \geq 1- 3\epsilon.
\end{equation}

For any~\(y \in C_Y \subset E_Y\) and~\(x \in A_y,\) we have by definition of~\(E_Y\) and~\(E_{XY}\) (see~(\ref{ey_def}) and~(\ref{exy_def})) that
\begin{eqnarray}\label{pxy_est}
p_{U \mid V}(x \mid y) &=& \frac{p_{UV}(x,y)}{p_V(y)} \nonumber\\
&\leq& \frac{2^{-n(H_{XY} -\epsilon)}}{2^{-n(H_Y+\epsilon)}} \nonumber\\
&=& 2^{-n(H_{XY}-H_Y - 2\epsilon)}.
\end{eqnarray}
Thus~\[\epsilon \leq \mathbb{P}_y(A_y)  = \sum_{x \in A_y} p_{U \mid V}(x \mid y) \leq \#A_y \cdot  2^{-n(H_{XY}-H_Y - 2\epsilon)} \] and so
we get that~\(\#A_y \geq \epsilon \cdot 2^{n(H_{XY}-H_Y-2\epsilon)}\) for each~\(y \in C_Y.\)

Suppose now that~\({\cal C}\) is any deterministic~\(n-\)length code of size at most~\(2^{n(H_{XY}-H_Y - 3\epsilon)}\) and let~\(f\) be any~\(Y-\)dependent encoding scheme based on~\({\cal C}\) as in Definition~\ref{def_one}. As in~(\ref{py_def}), we have for each~\(y \in C_Y\) that
\begin{eqnarray}\label{dulp}
\mathbb{P}_y\left(f_y({\cal C})\right) &=& \sum_{x \in f_y({\cal C})} p_{U \mid V}p(x \mid y) \nonumber\\
&\leq& 2^{n(H_{XY}-H_Y - 3\epsilon)} \cdot 2^{-n(H_{XY}-H_Y - 2\epsilon)} \nonumber\\
&=& \frac{1}{2^{n\epsilon}}
\end{eqnarray}
where the inequality in~(\ref{dulp}) follows from~(\ref{pxy_est}). Therefore
\begin{eqnarray}\label{uv_da}
\mathbb{P}\left(U \in f_V({\cal C})\right) &\leq& \mathbb{P}(V \notin C_Y) + \sum_{y \in C_Y}p(y) \frac{1}{2^{n\epsilon}} \nonumber\\
&\leq& 3\epsilon + \frac{1}{2^{n\epsilon}} \nonumber\\
&\leq& 4\epsilon
\end{eqnarray}
for all~\(n\) large. Thus the error probability~\(q(f,{\cal C})\) for any encoding scheme~\((f,{\cal C})\) is at least~\(1-4\epsilon\) for all~\(n\) large. This obtains the lower bound in~(\ref{err_low}) and therefore completes the proof of the Theorem.~\(\qed\)

\subsection*{Data Storage With Feature Selection}
Let~\(Z = (Z(1),\ldots,Z(m))\) be a random element chosen from some set~\({\cal Z}.\) We refer to the indices~\(i=1,2,\ldots,m\) as features and use statistical tests~\cite{kuhn} to determine whether~\(Z(i)\) and~\(Z(j)\) are correlated and obtain the dependency graph~\(G_{dep}.\) The vertex set of~\(G_{dep}\) is~\(\{1,2,\ldots,m\}\) and an edge with endvertices~\(i\) and~\(j\) belongs to~\(G_{dep}\) if and only if~\(Z(i)\) and~\(Z(j)\) are determined to be \emph{correlated}.

The rough idea behind the dependency graph is that if~\(i\) and~\(j\) are neighbours in~\(G_{dep},\) then~\(Z(i)\) and~\(Z(j)\) are highly correlated and we should lose little information if we were to discard either feature~\(i\) or feature~\(j.\)  In effect, given~\(G_{dep}\) and an integer~\(1 \leq d \leq m,\) we would like to obtain a subset~\(S\) containing at most~\(d\) features that satisfy the following properties:\\
\((i)\) The features present in~\(S\) are nearly uncorrelated.\\
\((ii)\) Each feature not present in~\(S\) is highly correlated with some feature present in~\(S.\)

For simplicity assume that using standard statistical tests (like for e.g. filters or wrappers, see Chapter~\(19,\)~\cite{kuhn}), we have determined~\(S = \{1,2,\ldots,d\}\) to be ``best" feature set and set~\(X := (Z(1),\ldots,Z(d))\) and~\(Y = (Z(d+1),\ldots,Z(m)).\) Using information from both~\(X\) and~\(Y,\) we would now like to find an encoding scheme that allows us to store~\(X\) using as few bits as possible. Indeed, suppose we have~\(n\) data points~\((X_i,Y_i), 1 \leq i \leq n\) and~(\ref{ent_cond}) holds together with~\(\frac{1}{n}\sum_{i=1}^{n}H(X_i) \longrightarrow H_X\) for some~\(H_X > 0.\) Applying Theorem~\ref{main_thm} with~\({\cal Y} = \emptyset\) we see that the relevant information~\((X_1,\ldots,X_n)\) can be stored using a binary code of length roughly~\(nH_X,\) with very small encoding error and without using any information from~\((Y_1,\ldots,Y_n).\)

On the other hand, if we use the ``side" information~\((Y_1,\ldots,Y_n),\) then we can store~\((X_1,\ldots,X_n)\) using a binary code of length roughly~\(n(H_{XY}-H_Y).\)  Using the fact that~\(H(X_i,Y_i) \leq H(X_i) + H(Y_i),\) we see that~\(H_{XY} \leq H_X+H_Y\) and so~\(1-\frac{H_{XY}-H_Y}{H_X}\) could be interpreted as the ``savings" obtained via dependent encoding resulting in ``smart" storage of data.

\setcounter{equation}{0}
\section{Neighbourhood Domination based Undersampling} \label{sec_neigh}
We begin with the definition and properties of neighbourhood domination in graphs and at the end of this section, explain its applications to data undersampling.

Let~\(K_n\) be the complete graph with vertex set~\(\{1,2,\ldots,n\}\) and let~\(H\) be any deterministic subgraph of~\(K_n.\) We say that~\({\cal T} \subset \{1,2,\ldots,n\}\) is a \emph{dominating} set of~\(H\) if each vertex~\(v\) is either present in~\({\cal T}\) or is adjacent in~\(H\) to some vertex~\(u \in {\cal T}.\) Let~\(N(v)\) be the set of neighbours of~\(v\) in the graph~\(H\) and let~\(d(v) = \#N(v).\) For~\(0< \theta <1,\) we say that~\({\cal T}\) is a~\(\theta-\)\emph{neighbourhood dominating} set of~\(H\) if for each vertex~\(v,\) there are at least~\(\theta d(v)\) vertices of~\(N(v)\) present in~\({\cal T}.\)



Let~\(M_n = M_n(\theta)\) be the minimum size  of a~\(\theta-\)neighbourhood dominating set of~\(H.\) In the following result below, we obtain an upper bound for~\(M_n\) with conditions on the minimum and maximum vertex  degree of~\(H\) and then show that the bound is essentially optimal, using random graphs. Formally, let~\(Y(f), f \in K_n\) be independent random variables indexed by the edge set of~\(K_n\) and having distribution
\[\mathbb{P}(Y(f)=1) = p = 1-\mathbb{P}(Y(f) =0)\] where~\(0 < p = p(n) < 1.\) Let~\(G\) be the homogenous random subgraph of~\(K_n\) formed by the set of all edges~\(e\) satisfying~\(Y(f)=1.\)

We have the following bounds for~\(M_n\) in terms of the degree parameters of the deterministic graph~\(H\) and the random graph realization~\(G.\)
\begin{theorem}\label{res_one} We have:\\
\((a)\) Let~\(\Delta\) and~\(\delta\) respectively denote the maximum and minimum vertex degree in~\(H\) and for~\(\eta > 0\) let~\(z := \frac{\eta^2(\theta+\eta)}{4}.\) Any~\(\theta-\)neighbourhood dominating set of~\(H\) has size at least~\(\frac{\theta \delta n}{\Delta^2}\) and  if
\begin{equation}\label{suff_cond}
4\Delta^2e^{-z\delta} \leq 1 \text{ and } e^{-z}  + 2e^{-z\delta} < 1
\end{equation}
strictly, then there is a~\(\theta-\)neighbourhood dominating set of size at most~\((\theta+2\eta)n.\)\\
\((b)\) For every~\(\epsilon > 0\) there are constants~\(M,D > 0\) such that if~\(p \geq \frac{M\log{n}}{n},\) then
\begin{equation}\label{low_bound}
\mathbb{P}\left((\theta-\epsilon)n \leq M_n \leq (\theta+\epsilon) n\right) \geq 1-e^{-Dnp}.
\end{equation}
\end{theorem}
In other words, with high probability, i.e. with probability converging to one as~\(n \rightarrow \infty,\)  we see that the size~\(M_n\) of a~\(\theta-\)neighbourhood dominating set is roughly of the order of~\(\theta n.\)

In the context of data analysis, the vertices of~\(K_n\) represent data points and for practical reasons explained at the end of this section, it is often important to undersample or reduce the data set size to a fraction~\(\theta n\) while allowing for enough neighbours per data point. Theorem~\ref{res_one} ensures that there exists such a set under the sufficient condition~(\ref{suff_cond}) which usually holds for most cases of interest as demonstrated at the section end.

In our proof of Theorem~\ref{res_one} below, we use the following results regarding the deviation estimates of sums of independent Bernoulli random variables and the Lov\'asz Local Lemma, which we state together as a Lemma for convenience.
\begin{lemma}\label{lemmax}
\((a)\) Let~\(\{W_j\}_{1 \leq j \leq r}\) be independent Bernoulli random variables satisfying~\(\mathbb{P}(W_j = 1) = 1-\mathbb{P}(W_j = 0) > 0.\) If~\(S_r := \sum_{j=1}^{r} W_j, \theta_r := \mathbb{E}S_r\) and~\(0 < \gamma \leq \frac{1}{2},\) then
\begin{equation}\label{conc_est_f}
\mathbb{P}\left(\left|S_r - \theta_r\right| \geq \theta_r \gamma \right) \leq 2\exp\left(-\frac{\gamma^2}{4}\theta_r\right)
\end{equation}
for all \(r \geq 1.\)\\
\((b)\) Let~\(A_1,\ldots,A_t\) be events in an arbitrary probability space. Let~\(\Gamma \) be the dependency graph for the events~\(\{A_i\},\) with vertex set~\(\{1,2,\ldots,t\}\) and edge set~\({\cal E};\) i.e. assume that each~\(A_i\) is independent of the family of events~\(A_j, (i,j) \notin {\cal E}.\) If there are reals~\(0 \leq y(i) < 1\) such that~\(\mathbb{P}(A_i) \leq y(i) \prod_{(i,j)\in {\cal E}} (1-y(j)),\) for each~\(i,\) then~\[\mathbb{P}\left(\bigcap_{i} A^c_i\right) \geq \prod_{1 \leq i \leq n} (1-y(i)) > 0.\]
\end{lemma}
For proofs of Lemma~\ref{lemmax}\((a)\) and~\((b),\) we refer respectively to Corollary A.1.14, pp. 312  and Lemma 5.1.1, pp. 64 of~\cite{alon}.

\emph{Proof of Theorem~\ref{res_one}\((a)\)}: Say that a set~\(V\) of vertices is~\(3-\)far if the graph distance (i.e. the number of  edges in the shortest path) between any two vertices is at least~\(3.\) For any vertex~\(v,\) there are at most~\(\Delta^2\) vertices at a distance~\(2\) from~\(v\) where~\(\Delta\) is the maximum vertex degree in~\(H.\) Therefore from Theorem~\(3.2.1,\) pp.~\(27\) of~\cite{alon}, we know that the graph~\(H\) contains a~\(3-\)far set~\({\cal T}\) of size at least~\(\frac{n}{2\Delta^2}.\) For any two vertices~\(u,v \in {\cal T},\) the corresponding neighbourhoods are disjoint; i.e.~\(N(u) \cap N(v) = \emptyset\) and so~\[M_n \geq \sum_{v  \in {\cal T}} \theta d(v) \geq \frac{\theta \delta n}{2\Delta^2}.\]

For the upper bound on~\(M_n,\) we use the probabilistic method. Select each vertex of~\(H\) with probability~\(x,\) independent of other vertices as follows. Let~\(Z_j, 1 \leq j \leq n\) be independent and identically distributed (i.i.d.) Bernoulli random variables, with~\[\mathbb{P}_Z(Z_j = 1) = x = 1-\mathbb{P}_Z(Z_j=0)\] and let~\(S = \{v : Z_v=1\}\) be the random set of chosen vertices. From the standard deviation estimate~(\ref{conc_est_f}), we get that
\begin{equation}\label{s_size}
\mathbb{P}_Z\left(\#S \leq (x+\eta)n\right) \geq 1-\exp\left(-\frac{\eta^2}{4}xn\right),
\end{equation}
where~\(\#S\) is the cardinality of~\(S.\)

Our goal is to show that if~\(x\) and~\(\eta\) are chosen appropriately, then~\(S\) is a~\(\theta-\)neighbourhood dominating set with positive~\(\mathbb{P}_Z-\)probability. Let~\(N_Z(v)=\{u \sim v : Z_u=1\}\) be the set of all vertices adjacent to~\(v\) in~\(G,\) that are also present in~\(S\) and let~\(A_v\) be the event that~\(\#N_Z(v) \geq (x-\eta)d(v).\) From the standard deviation estimate~(\ref{conc_est_f}) we get that
\begin{equation}\label{nikki_two}
\mathbb{P}_Z(A_v) \geq 1-\exp\left(-\frac{\eta^2xd(v)}{4}\right) \geq 1-\alpha
\end{equation}
where~\(\alpha := \exp\left(-\frac{\eta^2x}{4}\delta \right)\) and~\(\delta\) is the minimum vertex degree in~\(G.\)

The events~\(A_v\) and~\(A_u\) are independent if the graph distance between~\(u\) and~\(v\) satisfies~\(d_G(u,v) \geq 3\) and so if~\(\Delta\) denotes the maximum vertex degree in~\(H,\) then each~\(A_u\) depends on at most~\(\Delta^2\) of the events in~\(\{A_v\},\) which we denote as~\({\cal E}(v).\) This allows us to use Lov\'asz Local Lemma in Lemma~\ref{lemmax}\((b)\) under the assumption that
\begin{equation}\label{nikki_4}
4\Delta^2\alpha \leq 1.
\end{equation}
Indeed, setting~\(y(v) = 2\alpha\) we see that
\begin{equation}\label{thmulp2}
y(v)\prod_{u \in {\cal E}(v)}(1-y(u)) = 2\alpha(1-2\alpha)^{\Delta^2} \geq 2\alpha(1-2\alpha \Delta^2) \geq \alpha \geq \mathbb{P}_Z(A_v^c)
\end{equation}
where the third relation in~(\ref{thmulp2}) is true by~(\ref{nikki_4}) and the final relation in~(\ref{thmulp}) follows from~(\ref{nikki_two}).

From Lemma~\ref{lemmax}\((b),\) we therefore see that~\(\mathbb{P}_Z\left(\bigcap_{v} A_v\right) \geq \left(1-2\alpha\right)^{n}.\) Choosing~\(x-\eta= \theta\) and combining with~(\ref{s_size}), we then obtain that~\(S\) is a~\(\theta-\)neighbourood dominating set of size at most~\((\theta +2\eta) n,\) with~\(\mathbb{P}_Z-\)probability at least~\[\left(1-2\alpha\right)^{n} - \exp\left(-\frac{\eta^2}{4}xn\right) > 0\]
if
\begin{equation}\label{nikki_5}
\exp\left(-\frac{\eta^2}{4}x\right)  \leq 1-2\alpha-y
\end{equation}
for some small constant~\(y > 0.\) The conditions~(\ref{nikki_4}) and~(\ref{nikki_5}) complete the proof  of part~\((a)\) of the Theorem.~\(\qed\)

\emph{Proof of Theorem~\ref{res_one}\((b)\)}: The neighbourhood~\(N(v)\) of a vertex~\(v\) is the set of all vertices adjacent to~\(v\) in~\(G\) and the degree of~\(v\) is defined as~\(d(v) = \#N(v),\) the cardinality of~\(N(v).\)

For the upper bound, we use the conditions in Theorem statement to show that~\(G\) satisfies both the conditions~(\ref{nikki_4}) and~(\ref{nikki_5}) with high probability, i.e. with probability converging to one as~\(n \rightarrow \infty.\) The expected degree of any vertex~\(v\) is~\((n-1)p\)  and so using the deviation estimate~(\ref{conc_est_f}), we see that~\[\mathbb{P}\left(d(v) \geq \frac{np}{2}\right) \geq 1-e^{-Cnp}\] for some constant~\(C > 0.\) Setting~\(E_{deg} := \bigcap_{v=1}^{n}\left\{d(v) \geq \frac{np}{2}\right\}\) and using the union bound, we then get that
\begin{equation}\label{nikki_ax2}
\mathbb{P}\left(E_{deg}\right) \geq 1-ne^{-Cnp} \geq 1-ne^{-CM\log{n}} \geq 1-\frac{1}{n},
\end{equation}
provided~\(M\) is chosen sufficiently large. We fix such an~\(M\) henceforth and see that if~\(E_{deg}\) occurs, then the minimum vertex degree in~\(G\) is at least~\(\frac{np}{2} \geq \frac{M\log{n}}{2}.\) For all~\(n\) large, this implies that~(\ref{nikki_5}) holds, provided we fix~\(y\) small enough. Also, since the maximum vertex degree~\(\Delta \leq n,\) we can choose~\(M\) larger if necessary and ensure that~(\ref{nikki_4}) also holds. In effect, we see that if~\(E_{deg}\) holds, then there is a~\(\theta-\)neighbourhood dominating set of~\(G\) containing at most~\((\theta+2\eta)n\) vertices and this obtains the upper bound in~(\ref{low_bound}).

We prove the lower bound as follows. For any vertex~\(v\) the expected degree~\(\mathbb{E}d(v) = (n-1)p\) and so using the standard deviation estimate~(\ref{conc_est_f}), we get for~\(\epsilon > 0\) that~\[\mathbb{P}\left(d(v) \geq np(1-\epsilon)\right) \geq 1-e^{-Cnp}\] for some constant~\(C > 0.\) Letting~\(E_{deg} := \bigcap_{v=1}^{n} \left\{d(v) \geq np(1-\epsilon)\right\},\) we then get from the union bound that
\begin{equation}\label{e_deg_est}
\mathbb{P}(E_{deg}) \geq 1-ne^{-C_1np}
\end{equation}
for some constant~\(C_1 > 0.\)

Next, let~\(S\) be any set containing~\(\zeta n\) vertices with~\(\zeta < \theta\) strictly and let~\(m(S)\) be the number of edges of~\(G\) having both  endvertices in~\(S.\) We know that~\[\frac{\zeta^2n^2 p}{2} \geq \mathbb{E}m(S)={\zeta n \choose 2}p \geq \frac{\zeta^2n^2p}{4}\] and so by the deviation estimate~(\ref{conc_est_f}), we get that
\begin{equation}\label{ms_temp}
\mathbb{P}\left(m(S) \leq \frac{\zeta^2n^2p}{2}\right) \geq 1-e^{-C_2n^2p}
\end{equation}
for some constant~\(C_2 > 0.\) The number of choices for~\(S\) is at most~\(2^{n}\) and so letting~\(E_{edge}\) be the event that~\(m(S) \leq \frac{\zeta^2 n^2p}{2}\) for each~\(S\) containing~\(\zeta n\) vertices, we get from the union bound and~(\ref{ms_temp}) that
\begin{equation}\label{e_edge_est}
\mathbb{P}(E_{edge}) \geq 1-2^{n}e^{-C_2n^2p}.
\end{equation}

We assume henceforth that~\(E_{deg} \cap E_{edge}\) occurs, which by the union bound~(\ref{e_deg_est}) and~(\ref{e_edge_est}) happens with probability
\begin{equation}\label{e_tot_est}
\mathbb{P}(E_{deg} \cap E_{edge}) \geq 1-ne^{C_1 np} - 2^{n}e^{-C_2 n^2p} \geq 1-e^{-C_3np}
\end{equation}
for some constant~\(C_3 > 0\) provided~\(np > C_4 \log{n}\) for some sufficiently large constant~\(C_4 > 0.\) Let~\(S\) be any set of~\(\zeta n\) vertices. Because~\(m(S) \leq \frac{\zeta^2n^2p}{2}\) and the sum of the vertex degrees equals twice the number of edges, there exists a vertex~\(z \in S\) that is adjacent to at most~\[\zeta np = \frac{\zeta}{1-\epsilon} np(1-\epsilon) \leq \frac{\zeta}{1-\epsilon} d(z)\] vertices of~\(S,\) since~\(E_{deg}\) also occurs. We can choose~\(\epsilon\) small enough so that~\(\frac{\zeta}{1-\epsilon} < \theta\) strictly and obtain that no set containing~\(\zeta n\) vertices is a~\(\theta-\)neighbourhood dominating set of~\(G.\) Thus~\(M_n \geq \zeta n\) and from~(\ref{e_tot_est}) we then get the lower bound in~(\ref{low_bound}). This completes the proof of the Theorem.~\(\qed\)

\subsection*{Data Undersampling}
In many data classification problems involving two classes, there is often a large imbalance in the size of the majority and minority classes and this adversely affects the classwise accuracy of common predictive models like~\(k-\)Nearest Neighbour~(\(kNN\)), Random Forest, Logistic Regression or Neural Networks~\cite{he}~\cite{liu}~\cite{fern}. Therefore it is important to undersample or reduce the size of the majority class and make it comparable to the size of the minority class. Many different undersampling methods have been proposed in literature like random undersampling, near miss, condensed neighbour etc. and for more on  undersampling techniques in data analysis, we refer to~\cite{sui},~\cite{more}.

In this paper, we use neighbourhood domination as an undersampling methodology to obtain a balanced data set of given size. Indeed, suppose we are given~\(n\) data points from the majority class and~\(\theta n\) data points from the minority class for some~\(0 < \theta < 1.\) Instead of randomly selecting~\(\theta n\) points from the majority class, we could extract a more ``representative" set with the additional assumption that there is a distance metric~\(d(u,v)\) between two majority class data points~\(u\) and~\(v.\)

Henceforth any vertex represents a  data point belonging to the majority class. For a vertex~\(v,\) let~\(N_k(v)\) be the set of~\(k-\)nearest neighbours of~\(v,\) according to the metric~\(d.\) Let~\(H\) be the directed graph obtained by representing the data points as vertices and drawing a directed edge from vertex~\(u\) to vertex~\(v\) if~\(v \in N_k(u).\) The same analysis as in Theorem~\ref{res_one}\((a)\) holds with~\(\delta = \Delta = k\) and so if~(\ref{suff_cond}) holds then there is a~\(\theta-\)neighbourhood dominating set of size at most~\((\theta+2\eta)n.\)

Setting~\(k = M\log{n}\) for some large enough constant~\(M,\) we see that~(\ref{suff_cond}) is satisfied and so there exists a~\(\theta-\)neighbourhood dominating set~\(S\) of size\\\((\theta+2\eta)n.\) The advantage of using~\(S\) is that any vertex~\(v\) still has at least~\(\theta k\) neighbours from the set of its~\(k-\)nearest neighbours and this would aid greatly while performing classification, say for example, using~\(kNN\) type rules.


\setcounter{equation}{0}
\section{Conclusion} \label{sec_conc}
In this paper we have used a probabilistic approach towards data storage with feature selection and undersampling. We first obtained a Slepian-Wolf type result for nonstationary data and thereby demonstrated savings in data storage obtained by using information from the discarded features. We then considered neighbourhood domination in random graphs and explained how our methodology could be used to perform constrained undersampling in imbalanced datasets.

For the future we plan to apply our undersampling method to real world datasets and study the resulting minority class accuracy. We also would like to develop practical encoding schemes that use information from the discarded features and therefore result in smart data storage.

\subsubsection*{Acknowledgements}
I thank Professors Rahul Roy, Thomas Mountford, Federico Camia, Alberto Gandolfi, Lasha Ephremidze and C. R. Subramanian for crucial comments that led to an improvement of the paper. I also thank IMSc and IISER Bhopal for my fellowships.

\bibliographystyle{plain}

\end{document}